\documentclass{jfm}

\usepackage{graphicx}
\usepackage{natbib}

\ifCUPmtlplainloaded \else
  \checkfont{eurm10}
  \iffontfound
    \IfFileExists{upmath.sty}
      {\typeout{^^JFound AMS Euler Roman fonts on the system,
                   using the 'upmath' package.^^J}%
       \usepackage{upmath}}
      {\typeout{^^JFound AMS Euler Roman fonts on the system, but you
                   dont seem to have the}%
       \typeout{'upmath' package installed. JFM.cls can take advantage
                 of these fonts,^^Jif you use 'upmath' package.^^J}%
       \providecommand\upi{\pi}%
      }
  \else
    \providecommand\upi{\pi}%
  \fi
\fi

\ifCUPmtlplainloaded \else
  \checkfont{msam10}
  \iffontfound
    \IfFileExists{amssymb.sty}
      {\typeout{^^JFound AMS Symbol fonts on the system, using the
                'amssymb' package.^^J}%
       \usepackage{amssymb}%
         \let\leq=\leqslant
         
      }{}
  \fi
\fi

\ifCUPmtlplainloaded \else
  \IfFileExists{amsbsy.sty}
    {\typeout{^^JFound the 'amsbsy' package on the system, using it.^^J}%
     \usepackage{amsbsy}}
    {\providecommand\boldsymbol[1]{\mbox{\boldmath $##1$}}}
\fi

\newsavebox{\astrutbox}
\sbox{\astrutbox}{\rule[-5pt]{0pt}{20pt}}

\usepackage{amsmath}
\renewcommand{\b}[1]{\boldsymbol{#1}} 

\newcommand{\ud}{{\rm d}} 

\title[Ultrarelativistic shocks]{Ultra-relativistic geometrical shock dynamics and
vorticity}

\author[J. Goodman and A. MacFadyen]%
{J\ls E\ls R\ls E\ls M\ls Y\ns G\ls O\ls O\ls D\ls M\ls A\ls N$^{1,2}$
\thanks{Present address: Princeton University Observatory, Princeton, NJ
08544, U.S.A. e-mail: {\tt jeremy@astro.princeton.edu}}\\
A\ls N\ls D\ls R\ls E\ls W\ns M\ls A\ls C\ls F\ls A\ls D\ls Y\ls E\ls N$^{1,3}$%
}

\affiliation{$^1$ Institute for Advanced Study, Princeton, NJ 08544
\\[\affilskip]
$^2$Department of Astrophysical Sciences, Princeton University, 
Princeton, NJ 08544.
$^3$Department of Physics, New York University, New York, NY 10003.
}

\pubyear{2006}
\volume{---}
\pagerange{---}
\date{?? and in revised form ??}

\begin{document}

\maketitle

\begin{abstract}
  Geometrical shock dynamics, also called CCW theory, yields
  approximate equations for shock propagation in which only the
  conditions at the shock appear explicitly; the post-shock flow is
  presumed approximately uniform and enters implicitly \emph{via} a
  Riemann invariant.  The nonrelativistic theory, formulated by G.~B.
  Whitham and others, matches many experimental results surprisingly
  well.  Motivated by astrophysical applications, we adapt the theory
  to ultra-relativistic shocks advancing into an ideal fluid whose
  pressure is negligible ahead of the shock, but one third of its
  proper energy density behind the shock.  Exact results are recovered
  for some self-similar cylindrical and spherical shocks with
  power-law pre-shock density profiles.  Comparison is made with
  numerical solutions of the full hydrodynamic equations.  We review
  relativistic vorticity and circulation.  In an ultrarelativistic
  ideal fluid, circulation can be defined so that it changes only at
  shocks, notwithstanding entropy gradients in smooth parts of the
  flow.

\end{abstract}

Gamma-ray-burst afterglows have spurred us to look into this problem.
These cosmologically distant events are believed to involve a shock
launched by the death of a massive star with initial Lorentz factor
$\Gamma_0 > 10^2$ relative to a pre-shock circumstellar wind (mass
density $\rho_0\propto r^{-2}$) or interstellar medium
($\rho_0\sim\mbox{constant}$): see \citet{vPKW00,piran05,meszaros06}
for reviews.  Light curves fluctuate strongly at early times, probably
because of unsteadiness in the source; later, brightness falls
approximately as a power law in time but often with undulations that
may be due to inhomogeneities ahead of the shock.  The observed
radiation appears to be synchrotron emission, which implies that
$\gtrsim 10^{-2}$ of the postshock energy density takes the form of
magnetic field and highly relativistic electrons.  Even after
compression by the shock, typical circumstellar or interstellar fields
would be many orders of magnitude too small.  Therefore, it is often
supposed that magnetic energy is created rapidly by plasma
instabilities at the shock front \citep{medvedev_loeb99}.  We wish to
explore whether the compressed preshock field might instead be
amplified gradually by macroscopic fluid turbulence.  The source of
the turbulence is supposed to be vorticity produced as the shock
passes over inhomogeneities in the ambient medium.  Astrophysical
applications, however, are deferred to a later paper.  Our purpose
here is to develop and test suitable theoretical tools: a relativistic
version of geometric shock dynamics (hereafter GSD); and,
independently, a suitable redefinition of relativistic vorticity that
leads to conservation of the circulation on any fluid contour that
does not cross a shock.

The elements of nonrelativistic GSD were developed in the 1950s
\citep{moeckel52,chester54,chisnell57,whitham57,whitham58,whitham59}.
\citet{whitham74} gives a pedagogical review, upon which we have
relied heavily.  The one-dimensional version of the theory gives a
functional or even algebraic relationship (rather than a partial
differential equation) between variations in the pre-shock density and
variations in the shock Mach number.  The multidimensional version
describes the effect of changes in shock area---divergence or
convergence of the shock normals---on the Mach number.  Thus, the
theory reduces the dimensionality of the problem by one: in three
dimensions, for example, it gives a closed set of equations for the
evolution of the shock surface.  
GSD has even been adapted to reacting flows 
\citep[detonation waves:][]{li_ben-dor98}.

Naturally, there is a price to be paid in accuracy for these
simplifications.  For a recent critique, see \citet{baskar_prasad05}.
Nevertheless, GSD often performs remarkably well when there is reason
to expect that fluid gradients or geometrical constraints near the
shock ought to dominate, rather than reflections from boundaries
behind the shock, and even in some cases where there is no such
expectation.  GSD successfully
describes diffraction of shocks around corners, acceleration of
converging shocks, and even the propagation of kinks (``shock
shocks'') along shock fronts, as judged by comparisons with experiment
and with exact self-similar solutions \citep[and references
therein]{bryson_gross61,schwendeman88,whitham74}.

Relativistic units in which the speed of light $c=1$ will be used. We
adopt the conventions of \citet{schutz90} for tensors; in particular,
the metric in Minkowski coordinates $x^\mu=(x^0=t,x^1,x^2,x^3)$ is
$\eta_{\mu\nu}=\mbox{diag}(-1,1,1,1)$, while $T^{\mu\nu}$ and
$T^{\bar\mu\bar\nu}$ denote the components of the same tensor in two
Lorentz frames $\mathcal{O},\;\mathcal{\bar O}$.
In all cases considered here, the energy density of the pre-shock fluid
will dominated by rest mass, so that pressure and turbulent motions can
be neglected ahead of the shock.

\section{Planar shocks}\label{sec:planar}

Our goal is to transcribe GSD for an ultrarelativistic ideal fluid.
Following \citet{whitham74}, we begin with the case that the
area of the shock is constant and the pre-shock density ($\rho_0$)
is stratified on planes parallel to the shock front.
In place of Mach number, we will be concerned with shock Lorentz
factor ($\Gamma$) or rapidity parameter ($\Phi$), the two being
related by $\Gamma\equiv\cosh\Phi\approx e^\Phi/2\gg 1$.  These
quantities are defined in the rest-frame of the pre-shock medium.

The construction of GSD proceeds in two parts.  First, the jump
conditions are derived from the basic conservation laws; these relate the post-shock
fluid properties to the pre-shock ones if $\Gamma$ is given.  This step is
potentially exact but simplifies after approximations based on $\Gamma\gg1$,
$\rho\approx 3P$, and $P_0\ll\rho_0$.
Next, characteristic equations are derived for the post-shock flow, and
the (uncontrolled) approximation is made that one of the Riemann invariants
has a known and uniform value behind the shock.

\subsection{Jump conditions}\label{subsec:jump}

For a planar shock propagating 
in the $x^1$ direction, the relevant
components of the energy-momentum tensor are
\begin{equation}\label{eq:EMbasic}
  T^{00}= (\rho+P)\gamma^2-P,\quad   T^{01}= (\rho+P)\gamma^2\beta,\quad
T^{11}= (\rho+P)\gamma^2\beta^2+P,
\end{equation}
where the fluid 4-velocity has components
\[
U^\mu\to(\gamma,\gamma\beta,0,0)\equiv(\cosh\phi,\sinh\phi,0,0)
\]
measured in the shock rest frame. The proper energy density $\rho$
and pressure $P$ are defined in the local $\emph{fluid}$ rest frame,
so that they are Lorentz invariants.

The jump conditions in the shock rest frame are that $T^{01}$ and $T^{11}$
should be continuous. Since $P\approx\rho/3$ behind the shock and
$P_0\ll\rho_0$ in front of it, these conditions become
\begin{equation}\label{eq:jump}
T^{01}:~~4P\gamma^2\beta= -\rho_0\,\Gamma^2\qquad
T^{11}:~~4P\gamma^2\beta^2 + P= \rho_0\,\Gamma^2.
\end{equation}
Consistent with the ultra-relativistic approximation, the pre-shock
three-velocity has been set to $-1$, which incurs an error
$\sim O(\Gamma^{-2})$.  Eliminating $\Gamma^2\rho_0$
between the two equations, dividing through by $P$, and setting
$\gamma^{-2}\to1-\beta^2$ yields $(3\beta+1)(\beta+1)=0$.  The root
$\beta=-1$ corresponds to no shock at all.  Therefore, $\beta=-1/3$ in
the shock frame.  In terms of the rapidity parameters of the fluid and
the shock, $\tanh(\Phi-\phi)=1/3$.   The latter is a covariant
formulation since a Lorentz boost along $x^1$ with velocity $v$ simply
adds $\tanh^{-1}v$ to both $\phi$ and $\Phi$.  Substituting
$\beta=-1/3$ and $\gamma^2=9/8$ into either of eqs.~(\ref{eq:jump}) yields
$P=(2/3)\Gamma^2\rho_0$. So, the jump conditions are
\begin{subequations}
\label{eq:jumpers}
\begin{eqnarray}
  \slabel{eq:jump_phi}
  \phi&=\Phi-\tanh^{-1}\frac{1}{3}=\Phi-\ln\sqrt{2}\,,\\[1ex]
  \slabel{eq:jump_zeta}
\zeta&\equiv\frac{\sqrt{3}}{4}\ln P
\approx\frac{\sqrt{3}}{4}\left(
\ln\rho_0+2\Phi-\ln6\right)\,.
\end{eqnarray}
\end{subequations}
The peculiar factor $\sqrt{3}/4$ will
simplify the characteristic equations below.

\subsection{Whitham's Characteristic Rule}

Equations (\ref{eq:jumpers}) give two relations among the four
variables $(\phi,\Phi,\zeta,\ln\rho_0)$, or equivalently,
$(\gamma,\Gamma,P,\rho_0)$.  They are exact up to terms
$O(\Gamma^{-2})$.  Whitham's formulation of GSD adds one more
condition: the Riemann invariant associated with the characteristics
that go upstream from the postshock flow toward the shock, $R_+$, is
supposed to have the same value as it would if the shock were a
transition between constant states.  The rationale is that the
perturbations to the shock front are supposed to be localized; they
are driven by small-scale density variations in the pre-shock fluid,
or by local geometrical constraints on the shock, and therefore these
perturbations are supposed to average out downstream.

So the next step is to derive the Riemann characteristics from the
equations of motion $T^{\mu\nu}_{~~\,,\nu}=0$.  This has been done
in greater generality elsewhere \citep{marti_muller}, but for
completeness we shall rederive the special case we need.
With eqs.~(\ref{eq:EMbasic}), the equations of motion become
\begin{eqnarray*}
T^{0\nu}_{~~\,,\nu}&\propto&(2\cosh2\phi+1)\frac{4}{\sqrt{3}}\dot\zeta
+(4\sinh2\phi)\dot\phi+(2\sinh2\phi)\frac{4}{\sqrt{3}}\zeta'
+(4\cosh2\phi)\phi'=0,\nonumber\\
T^{1\nu}_{~~\,,\nu}&\propto&(2\sinh2\phi)\frac{4}{\sqrt{3}}\dot\zeta
+(4\cosh2\phi)\dot\phi+(2\cosh2\phi-1)\frac{4}{\sqrt{3}}\zeta'
+(4\sinh2\phi)\phi'=0,
\end{eqnarray*}
in which the dots denote $\upartial/\upartial x^0$ and the primes $\upartial/
\upartial x^1$.
Rather than manipulate these equations directly, 
it is easier to boost into the local
rest frame where $\phi=0$,  find the characteristics there, and
then boost back.  Since $\cosh2\phi\to1$ and $\sinh2\phi\to0$, the
equations above reduce to
\[
\dot\zeta+\frac{1}{\sqrt{3}}\phi'=0,\qquad
\dot\phi+\frac{1}{\sqrt{3}}\,\zeta'=0.
\]
By adding and subtracting these, one sees that the characteristic
velocities are $\pm 1/\sqrt{3}$, and the corresponding invariants
$\zeta\pm\phi$.  Boosting along $x^1$ to any other frame simply adds
a constant to $\phi$. Therefore,
\begin{subeqnarray}
  \label{eq:Rplus}
R_+&\equiv\zeta+\phi~\mbox{is constant on}~C_+:~\left(\frac{dx}{dt}\right)_+
=\tanh\left(\phi+\tanh^{-1}\frac{1}{\sqrt{3}}\right),\\
  \label{eq:Rminus}
R_-&\equiv\zeta-\phi~\mbox{is constant on}~C_-:~\left(\frac{dx}{dt}\right)_-
=\tanh\left(\phi-\tanh^{-1}\frac{1}{\sqrt{3}}\right).
\end{subeqnarray}

In Whitham's approximation, $R_+$ is constant not only along the $C_+$
characteristics but everywhere in the postshock flow,
even immediately behind the shock. Therefore, evaluating $\phi$ and
$\zeta$ from the jump conditions (\ref{eq:jumpers}), one obtains
\begin{equation}
  \label{eq:lambda}
  \Phi+\lambda\,\ln\rho_0\approx\mbox{constant},\qquad
\mbox{where}~\lambda\equiv\sqrt{3}-\frac{3}{2}\approx0.232
\end{equation}
This approximate equation predicts how the shock speed slows in
response to a transitory increase in pre-shock density:
$\Gamma\propto\rho_0^{-\lambda}$.  With eqs.~(\ref{eq:jumpers}),
we get the corresponding changes in postshock rapidity
 and pressure:
\begin{equation}
  \label{eq:philam}
\phi+\lambda\,\ln\rho_0=\mbox{constant};\qquad
P\propto\rho_0^{1-2\lambda}.
\end{equation}

\subsection{Comparison with an exact self-similar solution}

To reiterate, the approximation (\ref{eq:lambda}) is intended to
describe localized and transitory fluctuations in the propagation of a
shock that has some prescribed average Lorentz factor $\bar\Gamma$ and
advances into a ``cold'' medium with some prescribed, but spatially
variable, pre-shock density $\rho_0$ and negligible internal motions
and pressure.  \citet{whitham74} shows that the original
nonrelativistic version of his theory approximates rather well the
self-similar propagation of a planar shock from $x<0$ into a power-law
density profile $\rho\propto(-x)^n$, even though this situation does
not entirely satisfy the assumptions of GSD.

The corresponding ultra-relativistic solution has been given by
\citet{Sari06a}.  In Sari's terminology, the case of interest is a
planar (dimensionality parameter $\alpha=0$) ``Type II'' shock with
density exponent $k=-n<0$.  The Type I (II) solutions are those in
which the powerlaw scaling of shock position with time can (cannot) be
deduced from global energy conservation.  Type II, where the scalings
are determined by local conditions near the shock---rather than the
inertia of the ``piston'' behind it---is the case for which one might
hope that Whitham's theory would have some success.  Indeed, Sari's
equation (26) shows that the shock evolves as
$\Gamma\propto\rho^{-\lambda}$ with $\lambda$ \emph{exactly} as in
eq.~(\ref{eq:lambda})!  It is not clear why the agreement should be
exact, but presumably the extreme simplicity of the ultrarelativistic
fluid equations is somehow responsible.

\section{Non-planar shocks}

No vorticity can be created by an exactly planar shock, yet the 1D
theory above may be adequate for estimating the vorticity produced
by encounters between an ultrarelativistic shock and a density
inhomogeneity.  Lorentz contraction
causes the lateral dimension of inhomogeneities viewed in the shock or
postshock frame to be larger by a factor $\Gamma\gg 1$ than the
longitudinal ones, so that changes in speed and pressure are impressed
upon the immediately postshock flow before it ``notices'' that the
changes differ at other lateral positions.  Thus it should usually
be sufficient to evaluate the flow changes from the 1D theory, and then
take lateral derivatives to evaluate the resulting vorticity.

Nevertheless, it is worthwhile to extend the ultrarelativistic version
of Whitham's theory to nonplanar shocks for several reasons:
\begin{itemize}
\item in order to study the stability of the shock;
\item in order to compare with exact spherical and cylindrical
  self-similar solutions, and with numerical tests such as refraction
around an (oblique) corner;
\item because the extension is not difficult.
\end{itemize}

The idea of Whitham's nonplanar extension is to insert a factor
representing changes in shock area into the conservative form of the
fluid equations.  Thus let $x^1$ be a coordinate measuring arc length
along the shock normal, and $x^2$ and $x^3$ be coordinates in the
shock surface defined in such a way that a point moving along the shock
normal maintains constant $(x^2,x^3)$.
The equations of motion are taken to be
\begin{align}\label{eq:areacons}
T^{00}_{~~~,\,0}+A^{-1}\left(A\,T^{01}\right)_{,\,1}&=0\nonumber\\
T^{10}_{~~~,\,0}+A^{-1}\left(A\,T^{11}\right)_{,\,1}&=0.
\end{align}
Here $A$ is the 2D Jacobian relating the area of an element of the
shock surface to its initial area.  More precisely, since the
characteristic equation does not hold across the shock, $A$ represents
the cross-sectional area of a bundle of streamlines immediately behind
the shock; since the pre-shock medium is assumed to be at rest, the
flow behind the shock is normal to it.

Equations~(\ref{eq:areacons}) are not fully equivalent to
$T^{\mu\nu}_{~~~;\,\nu}=0$: for $\mu=1$, they do not contain the part of
the covariant derivative associated with turning of the shock normal.
They do however represent the divergence or convergence of the normals,
which leads to area change and strengthens or weakens the shock.
Because of the terms involving $A$, the quantities $R_\pm$ are no longer
invariant along their respective characteristics $C_\pm$.
The equation for $R_+$ works out to
\begin{eqnarray}
  \label{eq:Rpluseqn}
  \left(\frac{\ud }{\ud s}\right)_+(\zeta+\phi) &=& -\frac{\sinh\phi}
{\sqrt{3}\,\sinh\phi+\cosh\phi}\left(\frac{\ud }{\ud s}\right)_+\ln A
\nonumber\\[1ex]
&\to& -\frac{1}{\sqrt{3}+1}\left(\frac{\ud }{\ud s}\right)_+\ln A\,,
\end{eqnarray}
where $(\ud/\ud s)_+\equiv\upartial_1+(v_+)^{-1}\upartial_1$ is the derivative along
the $C_+$ characteristic, and
$v_+\equiv\tanh[\phi+\tanh^{-1}(1/\sqrt{3})]$ is the characteristic
velocity.  The final form of eq.~(\ref{eq:Rpluseqn}) is in the
pre-shock rest frame where $\cosh\phi\approx\sinh\phi\approx 1$ up to
$O(\Gamma^{-2})$.

The jump conditions (\ref{eq:jumpers}) are unchanged.  Inserting these
into (\ref{eq:Rpluseqn}) yields
\begin{subequations}
  \label{eq:char3D}
\begin{align}
  &\frac{\ud }{\ud s}\left(\Phi+\lambda\ln\rho_0+\mu\ln A\right)\approx 0,\\
  & \lambda\equiv\sqrt{3}-\frac{3}{2}\,,\qquad \mu\equiv 3\sqrt{3}-5.
\end{align}
\end{subequations}
Following Whitham, we have made the approximation that the characteristic
equation applies on the shock, although its propagation speed 
[rapidity $\Phi=\phi+\tanh^{-1}(1/3)$] is not quite the same as
that of the characteristic [$\tanh^{-1}v_+=\phi+\tanh^{-1}(1/\sqrt{3})$]
Consistent with this approximation,
the derivative $\ud/\ud s$ in (\ref{eq:char3D}) is taken to be the derivative
with respect to arc length along the shock normal.  Equation (\ref{eq:char3D})
predicts that the shock decelerates locally where its area increases,
and accelerates where the area decreases.  This will tend to stabilize
corrugations in the shock front.


Equations (\ref{eq:char3D}) need to be supplemented by a vector equation for
the shock normals.
Introduce a function $\tau(x,y,z)$ such that the locus of the shock in 
Minkowski coordinates is described by $t=\tau(x,y,z)$.
The normal to the shock is then $\vec n= \vec\nabla\tau/|\vec\nabla\tau|$, and its
3-velocity is $\vec V= \vec n/|\vec\nabla\tau|$.
The area function $A$ of the shock satisfies
\begin{equation}
  \label{eq:ndiver}
  \vec\nabla\cdot\left(\frac{\vec n}{A}\right)=0.
\end{equation}
This is a purely geometrical, rather than dynamical, statement.
\citet{whitham74} demonstrates it by applying Gauss's Law to a ``flux tube''
whose sides are made up of integral curves of $\vec n$, and whose ends are elements
of the shock surface at different times.

\subsection{Comparison with nonplanar self-similar solutions}

\citet{Sari06a}'s equation (26) for Type II solutions is equivalent to
\begin{equation*}
\frac{\ud}{\ud r}\ln\Gamma= \alpha(5-3\sqrt{3})~-\left(\sqrt{3}-\frac{3}{2}
\right)\frac{\ud}{\ud r}\ln\rho,
\end{equation*}
$\rho\propto r^{-k}$ being the pre-shock density.  Here $\alpha=0,1,2$ for
planar, cylindrical, and spherical shocks, respectively.  Since
the area factor should scale as $A\propto r^\alpha$ in these three
geometries, equation (\ref{eq:char3D}) predicts Sari's result perfectly.

In these self-similar solutions, dimensionality plays a limited
role since individual shock normals are constant.  \citet{whitham74}
discusses applications of the nonrelativistic theory to
non-self-similar and truly multidimensional problems such as
refraction of shocks around corners and obstacles.  Here we can expect
eqs.~(\ref{eq:char3D} \& (\ref{eq:ndiver}) not to be exact since, as
noted above, they do not incorporate the full covariant derivatives in
the equations of motion, and since the streamlines behind the shock
are not perfectly straight.  

\subsection{Detailed treatment of initially planar shocks in two dimensions.}
These details will facilitate
comparison with numerical solutions of the full hydrodynamic equations for
two-dimensional test cases (\S\ref{sec:numerical}).  They also serve to
illustrate effects that involve changes in the shock normal,
including the transverse propagation of disturbances along the shock front.

We take $z$ to be the ignorable coordinate.
Following Whitham again,
let $\psi$ be the angle between the normal and the $x$ axis, so that
$\upartial\tau/\upartial x=|\vec\nabla\tau|\cos\psi=V^{-1}\cos\psi$ and
$\upartial\tau/\upartial y=V^{-1}\sin\psi$.  Equation (\ref{eq:ndiver}) can then
be rephrased as the two first-order equations
\begin{subequations}
    \label{eq:2shock}
  \begin{eqnarray}
    \frac{\upartial}{\upartial x}\left(V^{-1}\sin\psi\right)
   -\frac{\upartial}{\upartial y}\left(V^{-1}\cos\psi\right) &= 0\,,\\[1ex]
    \frac{\upartial}{\upartial x}\left(A^{-1}\cos\psi\right)
   +\frac{\upartial}{\upartial y}\left(A^{-1}\sin\psi\right) &= 0\,,
  \end{eqnarray}
\end{subequations}
of which the first is simply the statement that
$\upartial^2\tau/\upartial y\upartial x= \upartial^2\tau/\upartial
x\upartial y$.
%
Together with (\ref{eq:char3D}), equations (\ref{eq:2shock})
form a hyperbolic system.  This is especially clear in the paraxial
approximation where $\psi\sim O(\Gamma^{-2})$. To this order, we may then replace
$\sin\psi\to\psi$, $\cos\psi\to 1$, and $V\to 1-(2\Gamma^{2})^{-1}$, so that
eqs.~(\ref{eq:2shock}) become
\begin{subequations}
\label{eq:hyperbolic}
\begin{eqnarray}
\frac{\upartial\psi}{\partial x}+\frac{1}{2}\frac{\upartial}{\upartial y}\left(\psi^2-
\Gamma^{-2}\right)&=0,\\
\frac{\upartial}{\partial x}A^{-1}+\frac{\upartial}{\upartial y}\left(A^{-1}\psi\right)&=0.
\end{eqnarray}
\end{subequations}
With our ordering, the term in $\psi^2$ is of higher order than the
others---it results from taking $\cos\psi=1-\psi^2/2$ rather than
unity in the first of eqs.~(\ref{eq:hyperbolic})---but it does no harm and
in fact makes the characteristic velocities work out more neatly.

If the shock is initially planar and the pre-shock density initially
uniform, then (\ref{eq:char3D}) implies that
$\Phi+\lambda\ln\rho_0+\mu\ln A$ is constant throughout the flow.
Thus $\Gamma$ in the first of equations (\ref{eq:hyperbolic}) is to be regarded
as a function of $A$ and $(x,y)$, given by
\begin{equation}
  \label{eq:gammabar}
  \Gamma(A,x,y) = \bar\Gamma\bar\rho^{\lambda}\bar A^{\mu}\rho_0^{-\lambda} A^{-\mu},
\end{equation}
in which the barred quantities are constants pertaining to the initially uniform
medium and planar shock, and $\rho_0(x,y)$ is a prescribed function.
The characteristic velocities of the system (\ref{eq:hyperbolic})-(\ref{eq:gammabar}) are
\begin{equation}
  \label{eq:Vpm}
  \left(\frac{dy}{dx}\right)_{\pm}  = \psi\pm\mu^{1/2}\Gamma^{-1}.
\end{equation}
The factor of $\Gamma^{-1}$ is easy to interpret as a consequence of relativistic beaming.
A disturbance propagating at the speed of light along an otherwise planar shock front
would have transverse velocity $\ud y/\ud t = \pm\Gamma^{-1}$ in the pre-shock rest frame.
To leading order in $\Gamma^{-1}$, we may replace $\ud t$ with 
$\ud x$ in this expression, so with $\mu^{1/2}\approx 0.4429$, the characteristics
(\ref{eq:Vpm}) are subluminal.

Since eqs.~(\ref{eq:2shock}) and (\ref{eq:hyperbolic}) are in conservation form, we may use them to study discontinuities
in the shock front itself: ``shock shocks.''  For a shock shock propagating at slope
$U\equiv (dy/dx)_{\rm ss}$, the jump conditions  implied by (\ref{eq:hyperbolic}) are
\begin{subequations}
  \begin{eqnarray}
    \label{eq:parajump}
    \left[-2U\psi + \psi^2-\Gamma^{-2}\right] &=&0,\\
    \left[A^{-1}(\psi-U)\right] &=&0,
  \end{eqnarray}
\end{subequations}
where $[Q]$  denotes the discontinuity in quantity $Q$ across the shock shock.
Let us assume a homogenous pre-shock medium, $\rho_0=\bar\rho=$constant, $\Gamma=$constant.  Then
if $\psi=0$ and $A=1$ ahead of the shock shock, the post-shock-shock quantities $(\psi',A')$ satisfy
\begin{equation}
  \label{eq:postss}
  (U\Gamma)^2=\frac{(A')^{2\mu}-1}{(A')^2-1}\,,\qquad \psi'=U(1-A').
\end{equation}
Thus in the limit $A'\to 0$, we have $\psi'=U=1/\Gamma$, and it follows from
eq.~(\ref{eq:gammabar}) that $\Gamma'\to\infty$.  In the opposite limit
$A'\gg 1$---but still $A'\ll \Gamma^{1/\mu}$ so that $\Gamma'\gg1$ (else the ultrarelativisitic
approximation would not apply)---we have $\psi'=-(A')^\mu/\Gamma=-1/\Gamma'$ and $U=-\psi'/\Gamma'$.

Finally, because
gamma-ray-burst shocks are believed to emanate from effectively pointlike explosions,
it is of interest to consider a nearly spherical rather than planar shock.  This
case is effectively two-dimensional if the perturbations are axisymmetric.  We
take polar coordinates $(r,\theta,\phi)$ such that $\theta=0,\upi$ is the axis of
symmetry and define $\psi$ to be the angle between the normal and radial
directions, \emph{i.e.} $\vec n\cdot\vec r=\cos\psi$.  The analogs of eqs.~(\ref{eq:2shock})
then become
\begin{subequations}
  \label{eq:sphericalfull}
\begin{eqnarray}
\frac{\upartial}{\upartial r}\left(\frac{r\sin\psi}{V}\right) -
\frac{\upartial}{\upartial\theta}\left(\frac{\cos\psi}{V}\right) &=&0,\\[1ex]
\frac{\upartial}{\upartial r}\left(\frac{r^2\cos\psi}{A}\right) +
\frac{r}{\sin\theta}
\frac{\upartial}{\upartial\theta}\left(\frac{\sin\theta\sin\psi}{A}\right) &=&0,
\end{eqnarray}
\end{subequations}
and for $\psi\ll1$, $\Gamma\gg 1$, eqs.~(\ref{eq:hyperbolic}) become
\begin{subequations}
  \label{eq:sphericalpara}
\begin{eqnarray}
\frac{\upartial}{\upartial r}\left(r\psi\right) + \frac{1}{2}
\frac{\upartial}{\upartial\theta}\left(\psi^2-\Gamma^{-2}\right) &=&0,\\[1ex]
\frac{\upartial}{\upartial r}\left(\frac{r^2}{A}\right) +
\frac{r}{\sin\theta}
\frac{\upartial}{\upartial\theta}\left(\frac{\psi\sin\theta}{A}\right) &=&0,
\end{eqnarray}
while (\ref{eq:gammabar}) is unchanged.
\end{subequations}

\section{Relativistic vorticity}

This discussion in this section is independent of the approximations of
GSD, although the ultrarelativistic equation of state $P=\rho/3$ figures
prominently.

As explained above, we are motivated by the need to explain the amplification
of magnetic field behind the shocks associated with gamma-ray bursts, and
by the possible role of turbulence in this amplification.  Therefore,
it may be worthwhile to record our assumptions about the relation of
post-shock vorticity to the magnetic field.

It follows from the induction equation of ideal magnetohydrodynamics,
\begin{equation}\label{eq:induction}
\upartial_t\vec B=\vec\nabla\times(\vec v\times\vec B),
\end{equation}
that magnetic energy increases according to
\[
\frac{\ud }{\ud t}\int\vec B\cdot\vec B\,\ud^3\vec x
=\int B_i B_j \upartial_i v_j\,\ud^3\vec x\,,
\]
which involves the instantaneous shear ($\upartial_i v_j+\upartial_j
v_i-\frac{2}{3}\vec\nabla\cdot\vec v$) and convergence
($\vec\nabla\cdot\vec v$) of the velocity field rather than the
vorticity, which is its curl.  Nevertheless, vorticity is important to
secular amplification of the field by localized disturbances.  In an
ideal fluid, a localized \emph{nonvortical} disturbance evolves into
sound waves, whose oscillations produce only transitory changes in
magnetic energy, and which propagate away from their source.  Energy
in vortical motions, however, remains localized, and the shear between
neighboring eddies is expected to amplify the field exponentially on
their turnover time.

\subsection{Vorticity and circulation}

Non-relativistically, the vorticity
${\vec\omega}\equiv{\vec\nabla\times\vec v}$, where ${\vec v}$ is the
fluid three-velocity.  In a compressible but isentropic fluid without
shocks, Kelvin's Circulation theorem is
\begin{equation}
  \label{eq:circulation}
  \frac{\ud }{\ud t}\oint\limits_C \vec v\cdot d\vec l=0,
\end{equation}
where $C$ is closed contour advected by the flow.

The generalization of ${\vec\omega}$ and eq.~(\ref{eq:circulation}) to
relativistic flow is not entirely straightforward \citep{Eshraghi03}.
Let $\vec U=(U^0,U^1,U^2,U^3)$ be the 4-velocity of the fluid, so that
$\eta_{\mu\nu}U^\mu U^\nu=-1$.  In terms of the local rest-frame
energy density $\rho$ and pressure $P$, the energy-momentum tensor is
\begin{equation}
  \label{eq:emtensor}
  T^{\mu\nu}=(\rho+P)U^\mu U^\nu+\eta^{\mu\nu}P.
\end{equation}
The equations of motion 
\begin{equation}
  \label{eq:eom0}
T^{\mu\nu}_{~~,\,\nu}=0
\end{equation}
must be supplemented by an equation of state.  Normally this involves
two independent thermodynamic variables, \emph{e.g.}  $P=P(\rho,T)$,
$P(\rho,N)$, or $P(N,S)$, where $T$ is the rest-frame temperature, $N$
is the proper number density of conserved particles, and $S$ is the
entropy per particle.  An essential feature of ultra-relativistic
shocks is that the postshock particles are highly relativistic in the
fluid rest frame, so that $P=\rho/3$.\footnote{This assumes that the
  stress is isotropic, which is not at all obvious in astrophysical
  applications where the plasma is collisionless.}  To the extent that
the fluid is ideal [eq.~(\ref{eq:emtensor})], entropy and temperature
gradients then have no effect on the flow---except at
shocks, but even there they do not have to be addressed explicitly.

To illustrate this point, we consider a
general equation of state in which entropy \emph{does} influence the
dynamics.  The conservation of particle number is expressed by
\begin{equation}
  \label{eq:conserved}
  \left(NU^\mu\right)_{,\mu}=0.
\end{equation}
The vorticity turns out to be best formulated as
\begin{equation}
  \label{eq:vort}
  \Omega_{\mu\nu}\equiv -H_{\mu,\nu}+H_{\nu,\mu}\,,
\end{equation}
in terms of
the relativistic enthalpy $h$ and its associated current $H^\mu$:
\begin{equation}
  \label{eq:enthalpy}
  h\equiv\frac{\rho+P}{N}\,,\qquad   H^\mu\equiv hU^\mu\,.
\end{equation}

With the First Law $d(\rho/N)=TdS-Pd(1/N)$ in the form
$n^{-1}dP=dh-TdS$, eq.~(\ref{eq:eom0}) can be rewritten as
\begin{equation}
  \label{eq:eom}
  U^\nu H_{\mu,\nu}=T S_{,\,\mu}-h_{,\,\mu}\,.
\end{equation}
The vorticity equation then follows from the ``curl'' of this, namely
\begin{eqnarray}\label{eq:vortevol}
 -(U^\nu H_{\alpha,\nu})_{,\,\beta}+ (U^\nu H_{\beta,\nu})_{,\,\alpha}&=
U^\nu\Omega_{\alpha\beta,\,\nu}+U^\nu_{~,\,\beta}\Omega_{\alpha\nu}
+U^\nu_{~,\,\alpha}\Omega_{\nu\beta}\nonumber\\
&=T_{,\,\alpha}S_{,\,\beta}-T_{,\,\beta}S_{,\,\alpha}\,.
\end{eqnarray}
The last line above vanishes if the entropy is uniform,
$S_{,\,\mu}=0$, or more generally if there is only one independent
thermodynamic quantity so that $S=S(T)$.  The intermediate expression
is the Lie derivative $\mathcal{L}_{\vec U}$
of the vorticity considered as a 2-form,
\begin{equation*}
\b{\Omega}\equiv\b{\ud H}\equiv H_{\nu,\,\mu}\b{\ud x}^\mu\wedge
\b{\ud x}^\nu\,,
\end{equation*}
with respect to the 4-velocity $\vec U$.  The statement of
conservation of circulation is then
\[
\frac{\ud }{\ud \tau}\oint\limits_{\upartial A} H_{\alpha}\,dx^\alpha=
\mathcal{L}_{\vec U}\iint\limits_{A} \b{\Omega}=0\,,
\]
where $A$ is a surface advected by $\vec U$ and $\upartial A$ is the curve
bounding it, and $d/d\tau\equiv U^\mu\upartial_\mu$ is the convective
derivative.  Since
different parts of the fluid move with different Lorentz factors, the
surface $A$ and curve $\upartial A$ will not (in general) remain within
hyperplanes of constant Minkowski time $t$, unfortunately.

Now we specialize to the ultra-relativistic equation of state $P=\rho/3$.
The true entropy per particle is $S\propto\ln(P/N^{4/3})$, which will
not be uniform after the shock passes over density inhomogeneities.
However, if we define an ersatz number density $\tilde N\propto
P^{3/4}$ at some initial time, so that $P/\tilde N^{4/3}$ is spatially
uniform, and if we demand that $\tilde N$ evolve according to
(\ref{eq:conserved}) with $\tilde N$ instead of $N$, then $P/\tilde
N^{4/3}$ will remain uniform {\it in smooth parts of the flow}, though not
across shocks.  This follows because $U_\mu
T^{\mu\nu}_{~~~,\nu}=-3U^\nu P_{,\,\nu}-4PU^\mu_{~,\mu}=0,$ whence
$U^\nu(P/\tilde N^{4/3})_{,\nu}=0$.

We don't actually have to deal with $\tilde N$ directly.  Since 
$\tilde h\equiv (\rho+P)/\tilde N =4P^{1/4}$, we can simply
redefine the enthalpy current as 
\begin{equation}\label{eq:H1}
H^\mu=P^{1/4}U^\mu\qquad\mbox{(when $P=\rho/3$ only).}
\end{equation}
Then eq.~(\ref{eq:eom}) becomes
\begin{equation}
  \label{eq:eom1}
  U^\nu H_{\mu,\nu}= -\left(P^{1/4}\right)_{,\,\mu},\qquad
P=\left(H^\mu H_\mu\right)^2.
\end{equation}
The vorticity defined in terms of this $\vec H$ \emph{via}
(\ref{eq:vort}) is conserved in the sense that the righthand side of
(\ref{eq:vortevol}) vanishes in all smooth parts of the flow, even
after shocks.  In particular, if $\b{\Omega}=0$ initially then it
remains zero as long as the flow remains smooth.  The jump conditions
that follow from integrating (\ref{eq:eom0}) \emph{across} shocks are
not equivalent to the corresponding integral of (\ref{eq:vortevol}),
however, and so this vorticity can be created at shocks.

\subsection{Vorticity vector}
$\Omega_{\alpha\beta}=-\Omega_{\beta\alpha}$ would appear to have six
algebraically independent components.  In fact it has only three.
To see this, note that
\begin{eqnarray*}
-h_{,\,\mu}&=(H_\nu U^\nu)_{,\,\mu}=H_{\nu,\,\mu}U^\nu+H_\nu U^\nu_{~,\,\mu}
=H_{\nu,\,\mu}U^\nu+h(U_\nu U^\nu)_{~,\,\mu}
=H_{\nu,\,\mu}U^\nu.
\end{eqnarray*}
Therefore if the fluid is isentropic, or if it is
ultrarelativistic and the enthalpy current is defined by
eq.~(\ref{eq:H1}), then eq.~(\ref{eq:eom}) becomes
\begin{equation}
  \label{eq:noelectric}
U^\nu\Omega_{\mu\nu}=0.
\end{equation}
This means that in the local rest frame, where $U^\nu\to\delta^\nu_0$,
the ``electric'' components $\Omega_{i0}=-\Omega_{0i}$ of the
vorticity vanish, and only the three ``magnetic'' components
$\Omega_{ij}=-\Omega_{ji}$ survive.  
Let $\vec\omega$ be the three-vector field with components
$\omega_i=\epsilon_{ijk}\Omega_{jk}/2$ in an arbitrary
inertial frame (not necessarily coinciding with the local
fluid rest frame).  If one uses
eq.~(\ref{eq:noelectric})
to eliminate the inertial components $\Omega_{i0}$ from the identity
\begin{equation*}
  \Omega_{\alpha\beta,\gamma}+  \Omega_{\beta\gamma,\alpha}
+\Omega_{\gamma\alpha,\beta}=0,
\end{equation*}
which is the tensorial expression of $\b{\ud}(\b{\ud H})=0$,
the result is
\begin{equation}
  \label{eq:omegaeqn}
  \upartial_t\vec\omega-\vec\nabla\times(\vec v\times
\vec\omega)=0,
\end{equation}
where $\vec v$ is the three-velocity, $v^i=U^i/U^0$.  This is formally
identical to the nonrelativistic vorticity equation of an isentropic
fluid, except that $\vec\omega$ is $\vec\nabla\times\vec H$ rather
than $\vec\nabla\times\vec v$.  The derivation just given, which
parallels that of the induction equation
(\ref{eq:induction}), shows that
eq.~(\ref{eq:omegaeqn}) is relativistically covariant.

\section{Numerical simulations and tests}\label{sec:numerical}

In order to test the predictions of the GSD theory we have performed
a series of numerical simulations with the the RAM special
relativistic hydrodynamics code \citet{zhang_macfadyen06}.  The
simulations test the variations of pressure and Lorentz factor behind a
strong shock as it passes over density perturbations in the pre-shock
medium.  The simulations are performed on the domain $x=[0.0,1.0]$
with reflecting boundary at $x=0.0$ and outflow (zero gradient)
boundary at $x=1$ and adiabatic index $\gamma_a=4/3$.  In the notation
of the present paper, the equation of state is
$P=(\gamma_a-1)(\rho-mN)$, where $m$ is the rest mass per particle.

We have performed ten simulations in which the pressure is initially set to
$P=10^5$ for $x<0.001$ and $P=10^{-6}$ elsewhere, to create a
plane-parallel analog of an explosively driven spherical gamma-ray-burst
shock.
The rest-mass contribution $mN$ to the energy density for
$x=[0.8,0.9]$ is initially set to one of the ten perturbed values $\rho_p =
\{1.0, 1.2, 1.4, 1.6, 1.8, 2.0, 5.0, 10.0, 20.0, 30.0\}$ and to
$\rho_i=1.0$ elsewhere.  The ten simulations are otherwise identical.  A
strong relativistic shock with Lorentz factor $\Gamma \sim 10$ is initially
driven into the medium as the over-pressured region expands and a thin
relativistic shell forms behind it.  As the shock crosses the (positive)
density perturbation, its Lorentz factor decreases and post-shock pressure
increases with the values seen in Fig. 1.  Note that $\gamma$ is
the Lorentz factor of the fluid rather than that of the shock itself; for an
ultrarelativistic shock advancing into a cold, stationary medium, the jump
condition (\ref{eq:jump_phi}) implies that these two Lorentz factors are
represented simply by $\gamma=\Gamma/\sqrt{2}$, a relation that does not
depend upon the approximations of GSD.  We measure $\gamma$ and $P$ at
$t=0.9$, after the shock has crossed the perturbation region and is at
$x\approx 0.9$, and we record for that time slice the values $P_m$ and
$\gamma_m$ where the pressure reaches its maximum post-shock value.  In
Fig. 1 (left panel) we show the logarithms of $P_m$ (plus signs) and
$\gamma_m^{-1}$ (asterisks) versus $\log(\rho_p/\rho_i)$.  The ordinates have
been scaled by the values $P_\circ$ and $\gamma_\circ$ obtained from a
fiducial run with uniform density ($\rho_p=\rho_i=1$).  The lines show the
scaling predicted by the GSD approximation with slopes $\lambda = \sqrt{3} -
\frac{3}{2} \approx 0.232$ (solid) and $1-2\lambda \approx 0.536$ (dashed).
In Fig. 1 (right panel), we show the fractional difference between the GSD
prediction and the numerical simulation for maximum pressure (plus signs) and Lorentz
factor (asterisks).

We find that the GSD approximation works surprisingly well even for large
density perturbations, justifying its use up to density contrasts of factors
of ten or more.

\acknowledgments

We acknowledge the use of the Scheides cluster at the Institute for Advanced
Study.  The software used in this work was in part developed by the
DOE-supported ASCI/Alliance Center for Astrophysical Thermonuclear Flashes at
the University of Chicago.  

\begin{figure}
 \centering
 \leavevmode
 \columnwidth=.5\columnwidth
 \includegraphics[width={\columnwidth}]{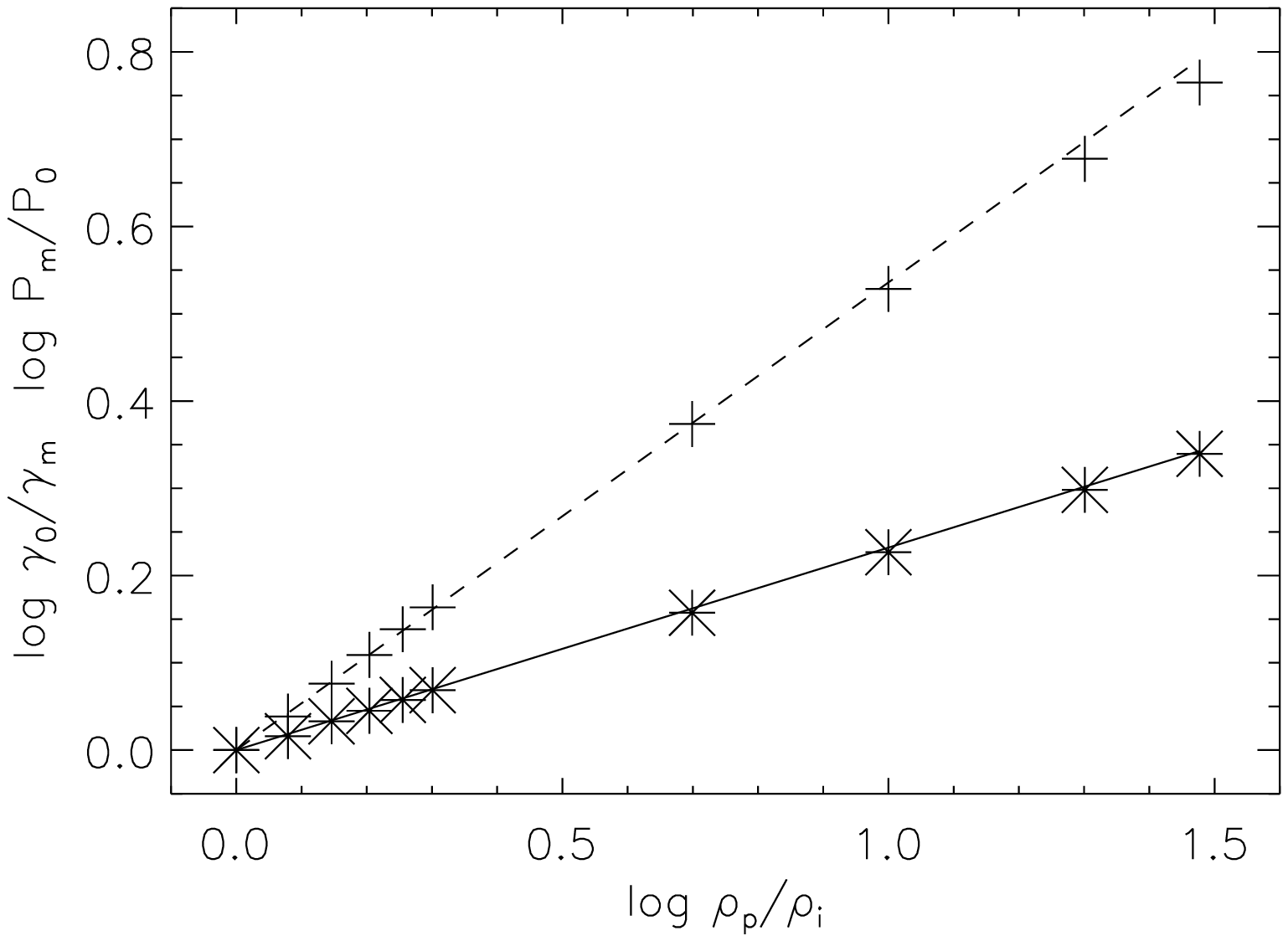}%
 \hfil
 \includegraphics[width={\columnwidth}]{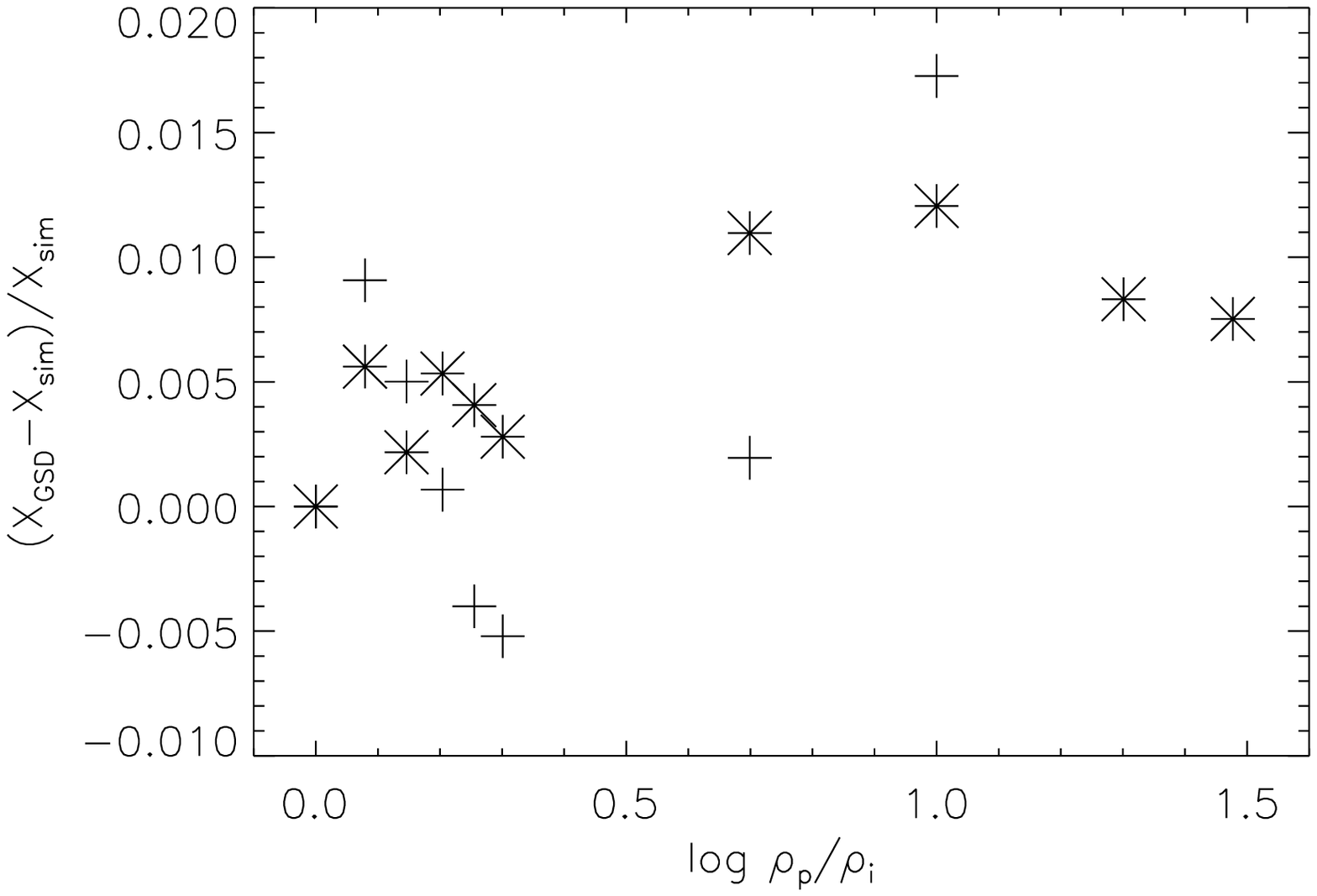}%
\caption{Comparison between analytic GSD predictions and numerical
simulations for post-shock Lorentz factor and pressure as a function
of the pre-shock density perturbation.  The right panel shows the fractional
difference between the GSD prediction $\mathrm{X_{GSD}}$ and the simulation
value $\mathrm{X_{sim}}$, where $\mathrm{X}$ stands for maximum pressure
$P_m$ (plus signs) or Lorentz factor $\gamma_m$ (asterisks).}
\end{figure}

\bibliography{ur}
\bibliographystyle{plainnat}

\end{document}